\documentclass[twocolumn,amssymb,aps]{revtex4}
\usepackage{amsmath}
\begin{document}
\title{Photon Momentum in Linear Dielectric Media}
\author{Michael E. Crenshaw}
\affiliation{Aviation and Missile Research, Development, and Engineering Center,\\
US Army RDECOM, Redstone Arsenal, AL 35898, USA}
\date{\today}
\begin{abstract}
According to the scientific literature, the momentum of a photon in a 
simple linear dielectric is either $\hbar\omega/(nc)$ or
$n\hbar\omega/c$ with a unit vector ${\bf \hat e}_k$ in the direction of
propagation.
These momentums are typically used to argue the century-old
Abraham--Minkowski controversy in which the momentum density of the
electromagnetic field in a dielectric is either the Abraham momentum
density, ${\bf g}_A={\bf E}\times{\bf H}/c$, or the Minkowski momentum
density, ${\bf g}_M={\bf D}\times{\bf B}/c$.
The elementary optical excitations, photons, are typically known as
polaritions in the particular case of light traveling in a dielectric
medium.
Applying the relativistic energy formula, we find that the total momentum
that is attributable to a polariton in a dielectric is
$\hbar\omega{\bf \hat e}_k/c$ corresponding to a total momentum
density ${\bf g}_T=n{\bf E}\times{\bf B}/c$.
\end{abstract}
\maketitle
\vskip 3.0cm
\par
In Maxwellian continuum electrodynamics, the speed of light in a simple
linear dielectric is $c/n$, where $n$ is the macroscopic refractive
index of the dielectric.
Adopting a more microscopic approach, Feynman \cite{BIFeynLect} argues
that the speed of light in a dielectric is always $c$, but only appears
to be $c/n$ due to interference between the source wave and reaction
fields that are generated by oscillating charges.
The scientific literature teaches us that a microscopic photon traveling
at speed $c$ in the vacuum slows to $c/n$ upon entering a
dielectric \cite{BIBarnLou,BIBarn}.
While the reduced velocity of the photon suggests that the momentum of
the photon becomes $\hbar\omega{\bf \hat e}_k/(nc)$, there are other
arguments for assigning the value $n\hbar\omega{\bf \hat e}_k/c$ to the
momentum of the photon \cite{BIBarnLou,BIBarn,BIPadg,BIGrzKemp,BIKemp},
where ${\bf\hat e}_k$ is a unit vector in the direction of propagation
of the macroscopic field.
The disparate momentums of a photon in a dielectric are often used as
proxies for the disputed momentum of the macroscopic electromagnetic
field in the Abraham--Minkowski momentum
debate \cite{BIBarnLou,BIBarn,BIKemp,BIPfei,BIBM,BIReview}.
In this scenario, the photon momentum
\begin{equation}
{\bf p}_A=\frac{\hbar\omega}{nc} {\bf\hat e}_k
\label{EQnp1.01}
\end{equation}
corresponds to the Abraham field momentum
\begin{equation}
{\bf G}_A=\int_{\sigma} \frac{{\bf E}\times{\bf H}}{c} dv 
\label{EQnp1.02}
\end{equation}
and 
\begin{equation}
{\bf p}_M=\frac{n\hbar\omega}{c} {\bf\hat e}_k
\label{EQnp1.03}
\end{equation}
is associated with the Minkowski field momentum
\begin{equation}
{\bf G}_M=\int_{\sigma} \frac{{\bf D}\times{\bf B}}{c} dv \, .
\label{EQnp1.04}
\end{equation}
The respective momentum densities are integrated over
all space $\sigma$.
The current consensus resolution of the Abraham--Minkowski controversy
is that both momentum formulas are correct \cite{BIBarn,BIPfei}.
\par
Photons are massless particles of electromagnetic energy that travel at
speed $c$ through the vacuum.
At the fundamental, microscopic level, dielectrics consist of tiny bits
of polarizable matter and host matter separated by relatively large
distances.
Then it might be argued that photons are massless particles of
electromagnetic energy that, between scattering events, travel at an
instantaneous speed $c$ through the interstitial vacuum of a rare
medium.
At the opposite, macroscopic, limit of matter, the refractive index $n$
is defined in terms of the effective speed of light $c/n$ in an
idealized model of dielectric matter that is continuous at all length
scales.
The basic excitations of a dielectric in the continuum limit are
polariton quasiparticles that contain energy of all the fields in the
dielectric \cite{BIHop,BIKnoesM}.
Such particles are spatially extensive and represent the total
macroscopic electromagnetic field over a macroscopic region of the
dielectric.
The total macroscopic field includes a polarization field that is
generated by charges in the material oscillating in reaction to the
source field.
The polarization reaction field travels with the propagating 
electromagnetic field.
\par
A polariton is identified with a specific quantity of electromagnetic
energy, a fact that requires the speed of a polariton to be the speed of
the electromagnetic field.
A microscopic description that details the instantaneous velocity of the
electromagnetic components of a polariton is outside the scope of
continuum electrodynamics; But, the effective speed of a polariton is
$c/n$ in accordance with the Feynman description of light propagation
in a dielectric \cite{BIFeynLect}.
In this work, we show that the effective momentum of the elementary
optical excitation of a dielectric is
\begin{equation}
{\bf p}=\frac{\hbar\omega}{c} {\bf\hat e}_k\,.
\label{EQnp1.05}
\end{equation}
The corresponding momentum of the macroscopic electromagnetic field,
\begin{equation}
{\bf G}_T=\int_{\sigma} \frac{n{\bf E}\times{\bf B}}{c} dv \, ,
\label{EQnp1.06}
\end{equation}
is the conserved total momentum in a thermodynamically closed system
that consists of a quasimonochromatic field incident on a
negligibly reflecting simple linear dielectric.
The total momentum, Eq.~(\ref{EQnp1.06}), has been proved to be
conserved for quasimonochromatic radiation incident on a
dilute, rare, or anti-reflection-coated simple linear dielectric
\cite{BIPfei,BIGord,BICrenBah,BIjmp}.
\par
Consider an inertial reference frame $S(\tau,x,y,z)$ with orthogonal
axes $x$, $y$, and $z$ and temporal coordinate $\tau$.
Position vectors are denoted by ${\bf x}=(x,y,z)$.
If a light pulse is emitted from the origin at time $\tau=0$, then
\begin{equation}
x^2+y^2+z^2-\left ( c\tau \right )^2=0 
\label{EQnp1.07}
\end{equation}
describes spherical wavefronts in the $S(\tau,x,y,z)$ system.
Writing time as a spatial coordinate $c\tau$, the
four-vector \cite{BIFinn1}
\begin{equation}
{\mathbb X}=(c\tau,{\bf x})=(c\tau,x,y,z)
\label{EQnp1.08}
\end{equation}
represents the position of a point in a dielectric-filled
four-dimensional flat non-Minkowski spacetime in which the temporal
coordinate is $\tau=t/n$.
Equation (\ref{EQnp1.08}) is a mathematically precise representation of
a point in the coordinate system $(c\tau,x,y,z)$.
Next, we want to investigate the physical implications of this
mathematical fact.
\par
Consider two inertial reference frames, $S(\tau,x,y,z)$ and
$S^{\prime}(\tau^{\prime},x^{\prime},y^{\prime},z^{\prime})$, in a
standard configuration \cite{BIRindler,BISchwarz} in which
$S^{\prime}$ translates at a constant velocity $u$ in the direction of
the positive $x$ axis and the origins of the two systems coincide at
time $\tau=\tau^{\prime}=0$.
If a light pulse is emitted from the common origin at time $\tau=0$,
then
\begin{equation}
(x^{\prime})^2+ (y^{\prime})^2+ (z^{\prime})^2
-\left ( c\tau^{\prime} \right )^2=0 
\label{EQnp1.09}
\end{equation}
describes wavefronts in the $S^{\prime}$ system and Eq.~(\ref{EQnp1.07})
holds for wavefronts in $S$.
It is relatively straightforward to derive transformations between these
coordinate systems by the usual methods of special
relativity \cite{BITipler}.
Now,
\begin{equation}
u=\frac{dx}{d\tau}= \frac{dx}{dt} \frac{dt}{d\tau}= vn \, .
\label{EQnp1.10}
\end{equation}
Then the transformation for $x$,
\begin{equation}
x=\gamma_d (x^{\prime}+u \tau^{\prime}) \, ,
\label{EQnp1.11}
\end{equation}
becomes
\begin{equation}
x=\gamma_d (x^{\prime}+nv \tau^{\prime}) \, .
\label{EQnp1.12}
\end{equation}
Similarly, the inverse transformation is
\begin{equation}
x^{\prime}=\gamma_d (x-nv \tau) \, .
\label{EQnp1.13}
\end{equation}
Substituting $x=c\tau$ and $x^{\prime}=c\tau^{\prime}$ into
Eqs.~(\ref{EQnp1.12}) and (\ref{EQnp1.13}), we eliminate
the temporal variables and obtain the material Lorentz
factor \cite{BIFinn1,BIRosen,BICrenuuu,BICrenSPIE}
\begin{equation}
\gamma_d=\frac{1}{\sqrt{1-\frac{n^2v^2}{c^2}}} \,.
\label{EQnp1.14}
\end{equation}
This derivation confirms the rather obvious phenomenological results
that are obtained by substituting the vacuum speed of light $c$ with
the speed of light $c/n$ in a dielectric \cite{BIRosen}.
This result differs from the usual Lorentz factor for a dielectric
\begin{equation}
\gamma=\frac{1}{\sqrt{1-\frac{v^2}{c^2}}} 
\label{EQnp1.15}
\end{equation}
that is derived using the relativistic velocity addition
theorem \cite{BIMoller} and is confirmed by the Fizeau frame-dragging
experiment.
No contradiction exists because the Fizeau experiment requires that
measurements are made in a vacuum-based laboratory reference frame 
\cite{BICrenuuu}.
Likewise, the velocity addition theorem is predicated on a laboratory
frame of reference for the observer.
For the physically distinct situation that we are considering here,
both inertial frames of reference reside within the dielectric
medium making Eq.~(\ref{EQnp1.14}) the correct Lorentz factor for our
system.
With further algebra, we obtain the complete material
Lorentz transformation \cite{BIFinn1}
\begin{subequations}
\label{EQnp1.16}
\begin{equation}
x=\gamma_d(x^{\prime}+nv\tau^{\prime})
\label{EQnp1.16a}
\end{equation}
\begin{equation}
y=y^{\prime}
\label{EQnp1.16b}
\end{equation}
\begin{equation}
z=z^{\prime}
\label{EQnp1.16c}
\end{equation}
\begin{equation}
\tau=\gamma_d\left (\tau^{\prime}+\frac{nv}{c^2}x^{\prime}\right ) \,.
\label{EQnp1.16d}
\end{equation}
\end{subequations}
for the case when both inertial reference systems are within the simple
linear dielectric.
\par
The invariant \cite{BIFinn1}
\begin{equation}
(\Delta \bar X_0)^2=(c/n)^2(\Delta t)^2-
((\Delta x)^2+(\Delta y)^2 +(\Delta z)^2) 
\label{EQnp1.17}
\end{equation}
can be written in terms of the spatial interval
\begin{equation}
c\Delta T =\frac{c}{n}\Delta t
\sqrt{1-\frac{n^2}{c^2} \left (
\left (\frac{\Delta x}{\Delta t}\right )^2+
\left (\frac{\Delta y}{\Delta t}\right )^2+
\left (\frac{\Delta z}{\Delta t}\right )^2 \right )}
\label{EQnp1.18}
\end{equation}
from which we obtain the interval of proper time
\begin{equation}
dT =\frac{dt}{\gamma_d n }.
\label{EQnp1.19}
\end{equation}
Taking the derivative of the position four-vector, Eq.~(\ref{EQnp1.08}),
with respect to the proper time, we obtain the four-velocity
\begin{equation}
{\mathbb U}=\frac{d{\mathbb X}}{dT}
=\frac{d{\mathbb X}}{dt}\frac{dt}{dT}
= \gamma_d n\left (
\frac{c}{n}, \frac{dx}{dt}, \frac{dy}{dt}, \frac{dz}{dt}
\right ) \, .
\label{EQnp1.20}
\end{equation}
The corresponding proper three-velocity is
\begin{equation}
{\bf u}=\gamma_d n {\bf v} \, . 
\label{EQnp1.21}
\end{equation}
Similarly, the four-momentum in a dielectric medium is
\begin{equation}
{\mathbb P}=m{\mathbb U}= 
\gamma_d n m \left ( 
\frac{c}{n}, \frac{dx}{dt},
\frac{dy}{dt},
\frac{dz}{dt}
\right )
\label{EQnp1.22}
\end{equation}
with a corresponding proper three-momentum
\begin{equation}
{\bf p}=\gamma_d n m {\bf v} \, .
\label{EQnp1.23}
\end{equation}
Substituting the material Lorentz factor, Eq.~(\ref{EQnp1.14}), into
the Einstein energy formula
\begin{equation}
E^2=M^2c^4=\gamma_d^2m^2c^4
=m^2c^4+(\gamma_d^2-1)m^2c^4
\label{EQnp1.24}
\end{equation}
yields
\begin{equation}
E^2=m^2c^4 +\gamma_d^2n^2v^2m^2c^2 \, .
\label{EQnp1.25}
\end{equation}
Substituting Eq.~(\ref{EQnp1.23}) into the previous equation, we find
\begin{equation}
E^2=m^2c^4+{\bf p}\cdot{\bf p} c^2
\label{EQnp1.26}
\end{equation}
for a dielectric medium.
Although the final result, Eq.~(\ref{EQnp1.26}), is identical to the
relativistic energy formula derived by Einstein for the vacuum,
several of the intermediate results, like the material Lorentz
factor, Eq.~(\ref{EQnp1.14}), the material Lorentz
transformation, Eq.~(\ref{EQnp1.16}), and the interval of proper
time, Eq.~(\ref{EQnp1.19}), are significant because they differ from
the well-known quantities that were derived in a different physical
setting.
\par
Polaritons are macroscopic compositions of the electromagnetic energy of
the microscopic source and reaction fields \cite{BIHop,BIKnoesM}.
Invoking Feynman \cite{BIFeynLect}, the microscopic fields that
contribute to the polariton travel at $c$, consequently, polaritons 
are required to be massless, just like photons.
The effective momentum of a massless particle of light in a dielectric
is given by Eq.~(\ref{EQnp1.26}) as 
\begin{equation}
{\bf p}=\frac{E}{c} {\bf\hat e}_k \, .
\label{EQnp1.27}
\end{equation}
Associating a volume $V$ with each polariton, we obtain the 
electromagnetic momentum density ${\bf g}={\bf p}/V$.
Integrating the momentum density over all-space we obtain the
momentum
\begin{equation}
{\bf G}_T
=\int_{\sigma} \frac{1}{c}\frac{E}{V} {\bf \hat e}_k dv
=\int_{\sigma} \frac{1}{2}\frac{n^2{\bf E}^2+{\bf B}^2}{c}
{\bf\hat e}_k \, dv \,.
\label{EQnp1.28}
\end{equation}
Then we can associate the momentum of a polariton with the total
momentum
\begin{equation}
{\bf G}_T
=\int_{\sigma} \frac{n {\bf E}\times{\bf B}}{c} dv
\label{EQnp1.29}
\end{equation}
by associating $|{\bf B}|$ with $n|{\bf E}|$ for quasimonochromatic
fields in the plane-wave limit.
Global conservation of the momentum quantity represented by
Eq.~(\ref{EQnp1.29}) is documented \cite{BIGord,BIPfei,BICrenBah} for
quasimonochromatic radiation incident on a dilute, rare, or
anti-reflection coated material.
As can be seen in Eq.~(\ref{EQnp1.28}), conservation of ${\bf G}_T$
is also guaranteed by conservation of electromagnetic energy.
We adopt current practice and define a polariton in terms of a fixed
amount of energy in an optical field of a given
frequency \cite{BIWhohas,BIAnd}
and substitute the Planck relation $E=\hbar\omega$ into
Eq.~(\ref{EQnp1.27}) to obtain
\begin{equation}
{\bf p}=\frac{\hbar\omega}{c} {\bf\hat e}_k
\label{EQnp1.30}
\end{equation}
for the effective momentum of a polariton.
\par
The result that is derived here differs from the prior art that is
displayed in Eqs.~(\ref{EQnp1.01}) and (\ref{EQnp1.03}).
The advantage of the current result is that the corresponding field
momentum, Eq.~(\ref{EQnp1.29}), is conserved for a quasimonochromatic
field incident on a negligibly reflecting stationary simple linear
dielectric.
The Abraham and Minkowski formulations assume a separate, material,
contribution to the momentum in order to preserve conservation of linear
momentum \cite{BIBarnLou,BIBarn,BIPadg,BIGrzKemp,BIKemp,BIPfei,BIBM}.
Assuming the Abraham form for the momentum of a polariton,
Eq~(\ref{EQnp1.01}), we spatially integrate the Abraham momentum 
density, ${\bf g}_A={\bf p}_A/V$, to obtain
\begin{equation}
{\bf G}_A
=\int_{\sigma} \frac{1}{c}\frac{E}{nV} {\bf \hat e}_k dv
=\int_{\sigma} \frac{1}{2}\frac{n^2{\bf E}^2+{\bf B}^2}{nc}
{\bf\hat e}_k \, dv \,.
\label{EQnp1.31}
\end{equation}
Associating $|{\bf B}|$ with $n|{\bf E}|$, as before, we obtain the
Abraham momentum formula
\begin{equation}
{\bf G}_A
=\int_{\sigma} \frac{ {\bf E}\times{\bf B}}{c} dv
\label{EQnp1.32}
\end{equation}
for the momentum of the macroscopic electromagnetic field.
For the thermodynamically closed system considered here, the total
momentum must be conserved.
By construction, the material kinetic momentum
\begin{equation}
{\bf G}_k = {\bf G}_T - {\bf G}_A =
\left (n-1 \right )
\int_{\sigma} \frac{ {\bf E}\times{\bf B}}{c} dv
\label{EQnp1.33}
\end{equation}
added to the Abraham momentum is equal to the total momentum.
Similarly, the material kinetic momentum that is associated with a
polariton is
\begin{equation}
{\bf p}_k=\frac{\hbar\omega}{c} {\bf\hat e}_k
-\frac{\hbar\omega}{nc} {\bf\hat e}_k
= \left ( 1-\frac{1}{n} \right ) \frac{\hbar\omega}{c} {\bf\hat e}_k \, .
\label{EQnp1.34}
\end{equation}
In this scenario, the effective mass of a polariton 
is
\begin{equation}
m_{eff}=\sqrt{E^2/c^4-({\bf p}_A
+{\bf p}_k)\cdot ({\bf p}_A+{\bf p}_k) /c^2 } =0,
\label{EQnp1.35}
\end{equation}
although the rest mass, which has an entirely different physical context,
continues to be $E/c=\hbar\omega/c$.
Turning to the Minkowski photon momentum, Eq.~(\ref{EQnp1.03}), we have
the material canonical momentum in terms of the macroscopic
electromagnetic field,
\begin{equation}
{\bf G}_c = {\bf G}_T - {\bf G}_M =
\left (n- n^2\right )
\int_{\sigma} \frac{ {\bf E}\times{\bf B}}{c} dv\, ,
\label{EQnp1.36}
\end{equation}
the material canonical momentum of a polariton,
\begin{equation}
{\bf p}_c=\frac{\hbar\omega}{c} {\bf\hat e}_k
-\frac{n\hbar\omega}{c} {\bf\hat e}_k
= \left ( 1-n \right ) \frac{\hbar\omega}{c} {\bf\hat e}_k \, ,
\label{EQnp1.37}
\end{equation}
and an effective mass of zero for the polariton.
\par
In conclusion, we derived the momentum of a polariton in a simple linear
dielectric that consists of an arbitrarily large region of space in
which the effective speed of light is $c/n$.
The consensus resolution of the Abraham--Minkowski controversy is that the
total momentum of the macroscopic field is composed of a two separate
components of momentum \cite{BIPfei}.
Historically, the total momentum was viewed as a composite of a field-only
momentum and a matter-only momentum where the question was the form of the
field component of momentum.
Recently, Barnett \cite{BIBarn} has shown that the total
momentum can viewed as a composite of the Minowski momentum and a material
canonical momentum, as well as the Abraham momentum and a material kinetic
momentum.
This is confirmed above as the two formulations are equivalent.
The advantage of the current formulation is that the theory can be cast
in terms of the total momentum without the necessity of separate handling
of the field and material parts.
\par

\end{document}